# Black phosphorus as saturable absorber for the Q-switched Er:ZBLAN fiber laser at 2.8 μm

Zhipeng Qin,[1] Guoqiang Xie,[1,*] Han Zhang,[2] Chujun Zhao,[3,*] Peng Yuan,[1] Shuangchun Wen,[3] and Liejia Qian[1]

[1] Key Laboratory for Laser Plasmas (Ministry of Education), Department of Physics and Astronomy, IFSA Collaborative Innovation Center, Shanghai Jiao Tong University, Shanghai 200240, China
[2] SZU-NUS Collaborative Innovation Center for Optoelectronic Science & Technology, Key Laboratory of Optoelectronic Devices and Systems of Ministry of Education and Guangdong Province, College of Optoelectronic Engineering, Shenzhen University, Shenzhen 518060, China
[3] Key Laboratory for Micro-/Nano-OptoelectronicDevices of Ministry of Education, School of Physics and Electronics, Hunan University, Changsha 410082, China
[*]xiegq@sjtu.edu.cn
[*]cjzhao@hnu.edu.cn

**Abstract**

Black phosphorus, a newly emerged two-dimensional material, has attracted wide attention as novel photonic material. Here, multi-layer black phosphorus is successfully fabricated by liquid phase exfoliation method. By employing black phosphorus as saturable absorber, we demonstrate a passively Q-switched Er-doped ZBLAN fiber laser at the wavelength of 2.8 μm. The modulation depth and saturation fluence of the black phosphorus saturable absorber are measured to be 15% and 9 μJ/cm$^2$, respectively. The Q-switched fiber laser delivers a maximum average power of 485 mW with corresponding pulse energy of 7.7 μJ and pulse width of 1.18 μs at repetition rate of 63 kHz. To the best of our knowledge, this is the first time to demonstrate that black phosphorus can realize Q-switching of 2.8-μm fiber laser. Our research results show that black phosphorus is a promising saturable absorber for mid-infrared pulsed lasers.

**Introduction**

Rapid progress has been made on two-dimensional (2D) materials represented by graphene, topological insulators (TIs) and transition metal dichalcogenides (TMDCs) in recent years [1-6]. Due to their broadband absorption, ultrafast carrier dynamics and planar characteristic [7-8], they have been regarded as the next-generation optoelectronics devices such as photoelectric detector, field-effect transistors, optical modulator, and so on [9-11]. So far, Q-switched and mode-locked lasers have been frequently reported with 2D materials as saturable absorber (SA) [12-17]. In the family of 2D materials, graphene is characterized by zero bandgap which makes it have extremely broadband optical response from visible to mid-infrared band [12]. However, its weak absorption results in a low modulation depth [15]. TIs are characterized by a full insulating gap in the bulk and gapless edge or surface states and TI SAs mainly work at 1, 1.5 and 2 μm wavelength at present [14,16,17]. TMDCs such as $MoS_2$ and $WS_2$ generally have large bandgap (1~2 eV) [13,18], which limits their application in the mid-infrared wavelength.

Black phosphorus (BP), a newly emerged 2D material, has gained wide attention recently. Up to now, it has been reported that BP is applied in sensor, field-effect transistors and solar cells [19-21]. Multi-layer BP has a similar structure with bulk graphite. In a single layer, each phosphorus atom is covalently bonded with three adjacent phosphorus atoms to form a stable honeycomb structure, and different layers are stacked together by van der Waals interaction [22]. Multi-layer BP has a direct energy bandgap structure, with bandgap from 0.3 eV to 2 eV depending on the number of layer [23]. Naturally, BP has the common properties of 2D materials such as wideband absorption, ultrafast carrier dynamics and planar characteristic [24]. The bandgap-controllable BP SAs can be fabricated by mechanical exfoliation method or liquid phase exfoliation (LPE) method [25, 26]. So far, the saturable absorption of BP has been demonstrated experimentally by Q-switched or mode-locked lasers from 0.6 to 2.0 μm wavelength. [27-30].

Here we experimentally demonstrate that BP SA is also feasible at the wavelength of 2.8 μm. The multi-layer BP, prepared by LPE method, was coated on a gold-film mirror as reflection-type saturable absorber mirror (SAM). By employing the fabricated BP SAM, we demonstrated a passively Q-switched Er-doped ZBLAN fiber laser at 2.8 μm. The Q-switched fiber laser delivered a maximum average power of 485 mW with corresponding pulse energy of 7.7 μJ and pulse width of 1.18 μs at repetition rate of 63 kHz. Due to the strong water absorption in body tissue for 3 μm laser, 3-μm pulsed laser are very useful in medical applications such as skin ablation, dentistry and cataract, etc.

**Preparation and characterization of BP SA**
In this work, the multi-layer BP was prepared by LPE method, which has been widely used to obtain 2D nanomaterials from layered bulk crystal. Firstly, we mixed bulk black phosphorous (30 mg) with N-Methyl pyrrolidone (NMP) solution (30 mL) together and sonicated at 40 kHz frequency and 300 W power for 10 hours for liquid exfoliation of bulk BP. Then, the supernatant liquor was obtained after centrifuging at 1500 rpm for 10 min. The as-prepared multi-layer BP flakes NMP solution (supernatant liquor) was dropped onto gold-film mirror as reflection-type saturable absorber mirror. After drying in cabinet, the BP SAM was used for characterization and laser experiments. The detailed characterization of morphology and Raman spectrum with same BP sample was performed in Ref [26]. The saturable absorption of the BP SAM was measured with a home-made mode-locked fiber laser at the wavelength of 2.8 μm, as shown in Fig. 1. By changing the incident fluence, the reflectivity of BP SAM increased from 79% to 91%. The measurement shows that the BP SAM has a modulation depth of 15% and saturation fluence of 9 $\mu J/cm^2$ at 2.8 μm.

**Experimental setup**
The schematic of the Q-switched fiber laser was shown in Fig. 2. The commercialized 976-nm laser diode (BTW, Beijing) was adopted as the pump source with maximum output power of 30 W, a core diameter of 105 μm and numerical aperture (NA) of 0.15. After collimated by a biconvex lens (F1=50 mm), the pump light was focused into first cladding configuration by the second biconvex lens (F2=100 mm). The 45° placed quartz mirror was antireflectively coated for pump light (T>95%) and highly reflectively coated for laser (R>99%). The double-cladding Er:ZBLAN fiber (FiberLabs, Japan) had a length of 4 m and Er-doping concentration of 6 mol.%. The core diameter of Er:ZBLAN fiber was 30 μm with NA of 0.12. The first cladding configuration had a diameter of 300 μm and NA of 0.5, which guaranteed efficient coupling of pump light. The pumping end facet of fiber was cut perpendicular to the fiber axis, with a Fresnel transmission of 96% as output coupler. At the tail end of fiber, it was cut with an angle of 8° to avoid parasitic oscillation. Then, two highly-reflective plane-convex mirrors (M1 and M2) with radii of curvature of 100 mm and 50 mm respectively, were used to reimage the end face of fiber onto BP SAM. The laser mode on the BP SAM was half of fiber core diameter.

**Experimental results and discussion**
With the setup of Fig.2, continuous-wave laser was generated at the threshold of incident pump power of 1.4 W. When the incident pump power increased to 2.2 W, the fiber laser started Q-switching operation. In the experiment the pulse train was captured by an infrared HgCdTe detector with a specified rise time of < 2 ns and working wavelength range of 1~9 μm (VIGO System model PCI-9), and displayed in a digital oscilloscope with 500-MHz bandwidth (Tektronix, DPO3054). The typical Q-switched pulse trains and pulse profiles were shown in Fig. 3 for different pump powers. At the Q-switching threshold, the fiber laser had an average output power of 145 mW, and pulse width of 2.1 μs and repetition rate of 39 kHz. The Q-switching operation can be maintained when the incident pump power increased continuously. At the incident pump power of 3.8 W, the average output power reached to 320 mW with a pulse width of 1.35 μs and repetition rate of 54 kHz. The shortest pulse width of 1.18 μs was obtained with an average output power of 485 mW and repetition rate of 63 kHz under an incident pump power of 5.4 W. The Q-switching operation turned to unstable when the incident pump power exceeded 5.4 W. It was worth noting that the position of BP-SAM was a key factor for Q-switching operation. In the experiment we carefully optimized the BP-SAM position for achieving the maximum output power and stable Q-switching operation.

Figure 4(a) shows the Q-switched average output power and pulse energy as a function of incident pump power. The average output power increased linearly from 145 mW to 485 mW with a slope efficiency of 10.6%. At the maximum output power of 485 mW, we obtained the maximum pulse energy of 7.7 μJ, which was higher than that generated from SESAM Q-switched fluoride fiber laser [31]. Up to now, it is also the maximum pulse energy generated from passively Q-switched 3 μm fiber lasers. Figure 4(b) shows the measured repetition rate and pulse width as a function of incident pump power. As expected, the repetition rate increased and pulse width decreased as the incident pump power increased. The repetition rate increased from 39 kHz to 63 kHz and pulse width decreased from 2.10 μs to 1.18 μs while the incident pump power varied from 2.2 W to 5.4 W.

Figure 5 shows the Q-switched pulse spectrum, which was measured by a mid-infrared spectral analyzer (Ocean Optics, SIR 5000) with a resolution of 0.22 nm. The spectral peak locates at 2779 nm with a FWHM of 4.6 nm.

**Conclusion**
In conclusion, multi-layer BP was fabricated by LPE method and 2.8 μm Q-switched fiber laser was experimentally demonstrated with BP as saturable absorber for the first time. The Q-switched fiber laser delivered a maximum average output power of 485 mW with pulse energy of 7.7 μJ, pulse width of 1.18 μs and repetition rate of 63 kHz. The BP is of low cost, easy fabrication, and variable bandgap, which makes it potential as a broadband saturable absorber for pulsed lasers, especially in the mid-infrared spectral regime where few saturable absorbers can work stably.
Acknowledgment

The work is partially supported by Shanghai Excellent Academic Leader Project (Grant No. 15XD1502100), National Natural Science Foundation of China (Grant No. 11421064) and National Basic Research Program of China (Grant No. 2013CBA01505).

**References**
1. K. S. Novoselov, A. K. Geim, S. V. Morozov, D. Jiang, Y. Zhang, S. V. Dubonos, I. V. Grigorieva, and V. V. Firsov, "Electric field effect in atomically thin carbon films," Science 306(5696), 666-669 (2004).
2. A. K. Geim, "Graphene: Status and prospects," Science 324(5934), 1530-1534 (2009).
3. J. E. Moore, "The birth of topological insulators," Nature 464(7286), 194-198 (2010).
4. H. Zhang, C. Liu, X. Qi, X. Dai, Z. Fang, and S. Zhang, "Topological insulators in Bi2Se3, Bi2Te3 and Sb2Te3 with a single dirac cone on the surface" Science 314(5806), 1757-1761 (2006).
5. A. A. Al-Hilli and B. L. Evans, "The preparation and properties of transiton metal dichalcogenide single crystals," J. Cryst. Growth 15(2), 93-101 (1972).
6. A. Anthony, E. Cobas, O. Ogundadegbe, and M. S. Fuhrer, "Realization and electrical characterization of ultrathin crystals of layered transition-meta dichalcogenides," J. Appl. Phys. 101, 014507 (2007).
7. K. F. Mak, M. Y. Sfeir, Y. Wu, C. H. Lui, J. A. Misewich, and T. F. Heinz, "Measurement of the optical conductivity of graphene," Phys. Rew. Lett. 101, 196405 (2008).
8. J. M. Dawlaty, S. Shivaraman, M. Chandrashekhar, F. Rana, and M. G. Spencer, "Measurement of ultrafast carrier dynamics in epitaxial graphene" Appl. Phys. Lett. 92, 042116 (2008).
9. T. J. Echtermeyer, L. Britnell, P. K. Jasnos, A. Lombardo, R. V. Gorbachev, A. N. Grigorenko, A. K. Geim, A. C. Ferrari, and K. S. Novoselov, "Strong plasmonic enhancement of photovoltage in graphene," Nat. Commun. 2, 458 (2011).
10. B. Radisavljevic, A. Radenovic, J. Brivio, V. Giacometti, and A. Kis, "Single-layer MoS2 transistors," Nat. Nanotechnol 6(3), 147-150 (2011).
11. M. Liu, X. Yin, E. Ulin-Avila, B. Geng, T. Zentgraf, L. Ju, F. Wang, and X. Zhang, "A graphene-based broadband optical modulator," Nature 474(7349), 64-67 (2011).
12. J. Ma, G. Xie, P. Lv, W. Gao, P. Yuan, L. Qian, U. Griebner, V. Petrov, H. Yu, H. Zhang, and J. Wang, "Wavelength-versatile graphemegraphem-gold film saturable absorber mirror for ultra-broadband mode-locking of bulk laser," Sci. Rep. 4, 5016 (2014).
13. Y. Zhang, S. Wang, H. Yu, H. Zhang, Y. Chen, L. Mei, A. D. Lieto, M. Tonelli, and J. Wang, "Atomic-layer molybdenum sulfide optical modulator for visible coherent light," Sci. Rep. 5, 11342 (2015).
14. C. Zhao, Y. Zou, Y. Chen, Z. Wang, S. Lu, H. Zhang, S. Wen, and D. Tang, "Wavelength-tunable picosecond soliton fiber laser with topological insulator:Bi2Se3 as a mode locker" Opt. Express 20(25), 27888-27895 (2012).
15. J. Ma, G. Q. Xie, P. Lv, W. L. Gao, P. Yuan, L. J. Qian, H. H. Yu, H. J. Zhang, J. Y. Wang, and

D. Y. Tang, "Graphene mode-locked femtosecond laser at 2 μm wavelength," Opt. Lett. 37(11), 2085-2087 (2012).
16. Z. Luo, Y. Huang, J. Weng, H. Cheng, Z. Lin, B. Xu, Z. Cai, and H. Xu, "1.06 μm Q-switched ytterbium-doped fiber laser using few-layer topological insulator Bi2Se3 as a saturable absorber," Opt. Express 21(24), 29516-29522 (2013).
17. Z. Luo, C. Liu, Y. Huang, D. Wu, J. Wu, H. Xu, Z. Cai, Z. Lin, L. Sun, and J. Weng, "Topological-insulator passively Q-switched double-clad fiber laser at 2 μm wavelength," IEEE J. Sel. Top. Quant. 20, 0902708 (2014).
18. Q. H. Wang, K. Kalantar-Zadeh, A. Kis, J. N. Coleman, and M. S. Strano, "Electronics and optoelectronics of two-dimensional transition metal dichalcogenides," Nat. nanotechnol 7(11), 699-712 (2012).
19. A. N. Abbas, B. Liu, L. Chen, Y. Ma, S. Cong, N. Aroonyadet, M. Kopf, T. Nilges, and C. Zhou, "Black phosphorus gas sensor," ACS Nano. 9(5), 5618-5624 (2015).
20. L. Li, Y. Yu, G. J. Ye, Q. Ge, X. Ou, H. Wu, D. Feng, X. H. Chen, and Y. Zhang, "Black phosphorus field-effect transistors," Nat. nanotechnol. 9(5), 372-377 (2014).
21. J. Dai and X. C. Zeng, "Bilayer phosphorene: effect of stacking order on bandgap and its potential applications in thin-film solar cells," J. Phys. Chem. Lett. 5(7), 1289-1293 (2014).
22. H. Liu, Y. Du, Y. Deng, and P. D. Ye, "Semiconducting black phosphorus: Synthesis, transport properties and electronic applications," Chem. Soc. Rev. 44(9), 2732-2743 (2015).
23. V. Tran, R. Soklaski, Y. Liang, and L. Yang, "Layer-controlled band gap and anisotropic excitions in few-layer black phosphorus," Phys. Rev. B 89, 235319 (2014).
24. J. Qiao, X. Kong, Z. Hu, F. Yang, and W. Ji, "High-mobility transport anisotropy and linear dichroism in few-layer black phosphorus," Nat. Commun. 5, 4475 (2014).
25. Y. Chen, G. Jiang, S. Chen, Z. Guo, X. Yu, C. Zhao, H. Zhang, Q. Bao, S. Wen, D. Tang, and D. Fan, "Mechanically exfoliated black phosphorus as a new saturable absorber for both Q-switching and mode-locking laser operation," Opt. Express 23(10), 12823-12833 (2015).
26. S. B. Lu, L. L. Miao, Z. N. Guo, X. Qi, C. J. Zhao, H. Zhang, S. C. Wen, D. Y. Tang, and D. Y. Fan, "Broadband nonlinear optical response in multilayer black phosphorus: an emerging infrared and mid-infrared optical material," Opt. Express 23(9), 11183-11194 (2015).
27. Z. Luo, M. Liu, Z. Guo, X. Jiang, A. Luo, C. Zhao, X. Yu, W. Xu, and H. zhang, "Microfiber-based few-layer black phosphorus saturable absorber for ultra-fast fiber laser," arXiv preprint, arXiv:1505.03035, 2015.
28. R. Zhang, Y. Zhang, H. Yu, H. Zhang, R. Yang, B. Yang, Z. Liu, and J. Wang, "Broadband black phosphorus optical modulator in visible to mid-infrared spectral range," arXiv preprint, arXiv:1505.05992, 2015.
29. D. Li, H. Jussila, L. Karvonen, G. Ye, H. Lipsanen, X. Chen, and Z. Sun, "Ultrafast pulse generation with black phosphorus," arXiv preprint, arXiv:1505.00480, 2015.
30. T. Jiang, K. Yin, X. Zheng, H. Yu, and X. Cheng, "Black phosphorus as a new broadband saturable absorber for infrared passively Q-switched fiber laser," arXiv preprint, arXiv:1504.07341.
31. J. F. Li, H. Y. Luo, Y. L. He, Y, Liu, L. Zhang, K. M. Zhou, A. G. Rozhin, and S. K. Turistyn, "Semiconductor saturable absorber mirror passively Q-switched 2.97 μm fluoride fiber laser," Laser Phys. Lett. 11, 065102 (2014).

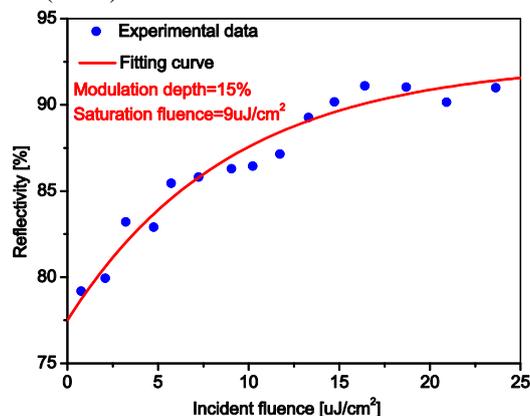

Fig. 1. The saturable absorption measurement of BP SAM at 2.8 μm wavelength.

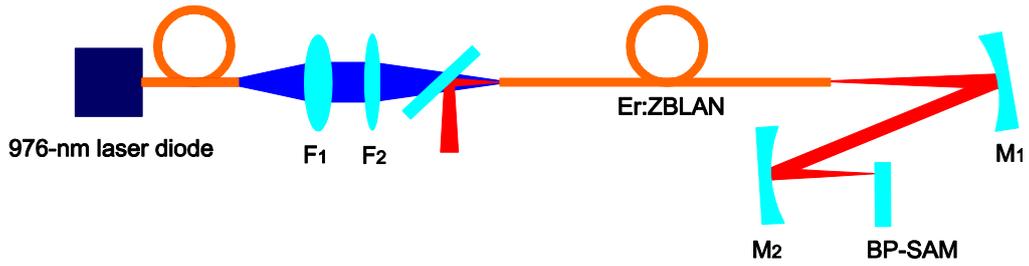

Fig. 2. The schematic of the passively Q-switched Er:ZBLAN fiber laser. BP-SAM, black phosphorus saturable absorber mirror.

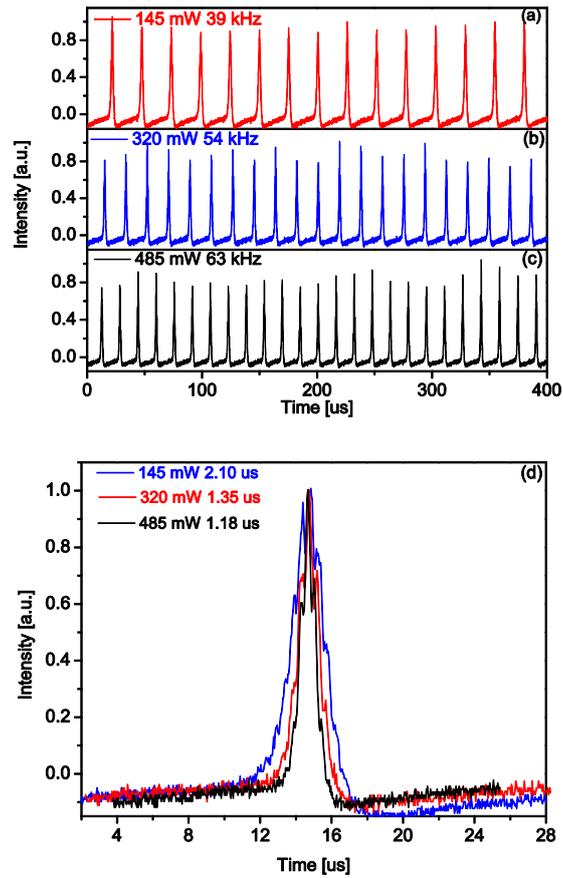

Fig. 3. (a-c) Q-switched pulse trains at the output powers of 145 mW, 320 mW and 485 mW, respectively. (d) Their corresponding pulse profiles.

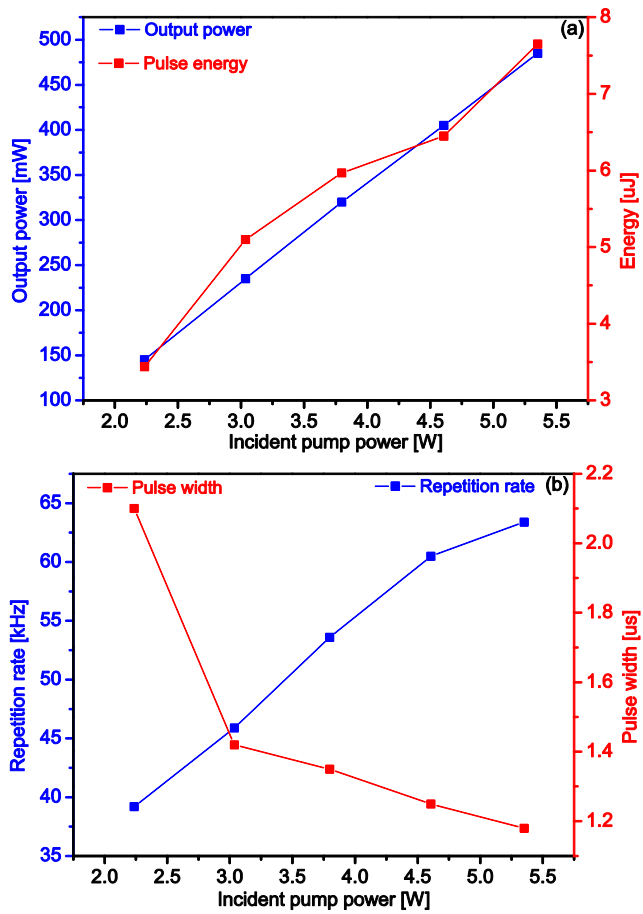

Fig. 4. (a) Average output power and pulse energy, (b) repetition rate and pulse width as a function of incident pump power.

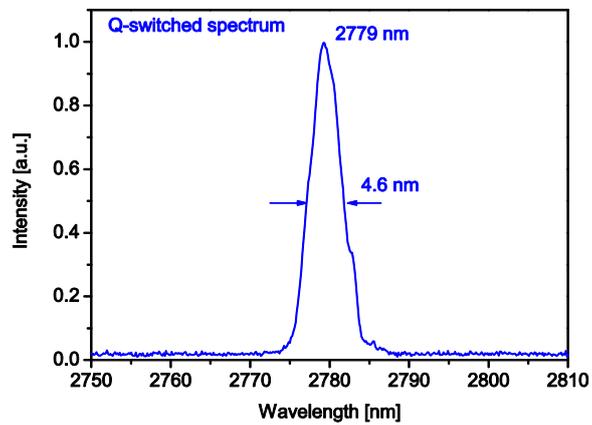

Fig. 5. The Q-switched pulse spectrum measured at the maximum output power.